\documentclass[
aps,%
11pt,%
final,%
notitlepage,%
oneside,%
onecolumn,%
nobibnotes,%
nofootinbib,%
superscriptaddress,%
noshowpacs,%
centertags]%
{revtex4}

\usepackage{multirow}

\begin{document}

\title{Kinematics and Stellar Disk Modeling of  Lenticular
Galaxies}

\author{\firstname{A.~V.}~\surname{Zasov}}
\affiliation{Sternberg State Astronomical Institute, Moscow State
University, Moscow, 119992 Russia}

\author{\firstname{A.~V.}~\surname{Khoperskov}}
\affiliation{Volgograd State University, Volgograd, 400062 Russia}

\author{\firstname{I.~Yu.}~\surname{Katkov}}
\affiliation{Sternberg State Astronomical Institute, Moscow State
University, Moscow, 119992 Russia}

\author{\firstname{V.~L.}~\surname{Afanasiev}}
\affiliation{\saoname}

\author{\firstname{S.~S.}~\surname{Kaisin}}
\affiliation{\saoname}

\received{February 17, 2012}  \revised{July 2, 2012}

\begin{abstract}
We present the results of spectroscopic observations of three S0--Sa
galaxies: NGC\,338, NGC\,3245, and NGC\,5440 at the SAO RAS 6-m BTA
telescope. The radial distributions of the line-of-sight velocities
and radial velocity dispersions of stars and ionized gas were
obtained, and rotation curves of galaxies were computed. We
construct the numerical dynamic  $N$-body galaxy models with $N \geq
10^6$ point masses. The models include three components: a ``live''
bulge, a collisionless disk, dynamically evolving to the marginally
stable state, and a pseudo-isothermal dark halo. The estimates of
radial velocities and velocity dispersions of stars obtained from
observations are compared with model estimates, projected onto the
line of sight. We show that the disks of NGC\,5440 and the outer
regions of NGC\,338 are dynamically overheated. Taking into account
the previously obtained observations, we conclude that the dynamic
heating of the disk is present in a large number of early-type disk
galaxies, and it seems to ensue from the external effects. The
estimates of the disk mass and  relative mass of the dark halo are
given for seven galaxies, observed at the BTA.
\end{abstract}

\maketitle

\section{INTRODUCTION}

Lenticular galaxies (S0) are disk systems which generally differ
from spiral galaxies by a very small amount of neutral gas and young
stars in the disk. As a consequence, they do not have any contrast
spiral arms or vigorous centers of star formation. These galaxies
exist not only in clusters and groups of galaxies, but also among
the field galaxies. Alike the spiral galaxies, the
 luminosity ratio of disk and bulge in  lenticular galaxies varies
in a broad range of values, although on the average it is higher.
In most cases, similar to the spirals, the bulges here are
actually the pseudo-bulges~\cite{Noord07}. The surface
brightnesses of the disks of S0 galaxies are not lower but rather
higher than those in the spiral galaxies of similar
luminosity~\cite{Boselli06, Boselli09}. At the
same time, there is no abrupt transition from the S0 galaxies to
the later types (S0/a, Sa).

The question about the features of S0 galaxies and their
evolutionary relationship with the spiral galaxies is actively
discussed in the literature
(see~\cite{Boselli09,Barr07,Aragon07,Bekki11,vdBergh09,Williams10,Sil-Chil11,Noord08}
and references therein). Various authors have considered several
different scenarios of the transformation of a galaxy with an
initially intensive star formation in the disk into a lenticular
galaxy. They may be linked both with the internal causes (high
efficiency of star formation leading to the depletion of gas, the
sweeping-out of gas from the disk at the stage of active nucleus or
a burst of star formation) and external (heating up or sweeping-out
of gas by the interaction with the intergalactic medium, an
cessation of accretion into the disk,  merging of small galaxies,
followed by a burst of star formation, tidal interactions with
nearby galaxies or a massive cluster of galaxies). Merging with
smaller systems and strong tidal perturbations, even if they
occurred in the distant past, leave their nonvanishing trace,
dynamically heating the stellar disk~\mbox{\cite{Bekki11,
Bournaud05}}. At the same time though, the sweeping-out of
gas or its rapid depletion at the epoch when the disk has already
been mostly formed, would not significantly affect its dynamics.

The observations show that lenticular galaxies significantly differ
not only by the shapes of their rotation curves, but also by the
radial profiles of velocity dispersion of their stellar components.
Velocity dispersion, measured along the line of sight, is typically
decreasing monotonously with distance from the center. But in some
cases, even at a distance of two radial scales from the center of
the disk ($r=2h$), the radial component of dispersion would set at a
level of one-half of the rotation velocity (the corresponding
estimates, based on the published data are given in
Table~\ref{tab1_log_obs} from Zasov et
al.~\cite{ZKhS11}). In other cases, the velocity dispersion
turns out to be the same as for most of spiral galaxies. The
kinematic models show that the kinematics of the disks of some
lenticular galaxies does not demonstrate any anomalies. As an
example, note the measurements of the kinematics of the planetary
nebulae in the disk of NGC\,1023 by Cortesi et
al.~\cite{Cortesi11}. In such cases, we can conclude that
the evolution of a given lenticular galaxy has progressed in a
relaxed manner, without any considerable external perturbations.

Obviously, neither the magnitude of the velocity dispersion, nor its
relation to the rotation velocity would already imply whether the
galaxy has experienced a strong heatup of the dynamic disk, or the
velocity dispersion of the disk stars is close to the minimum
(marginal) level, ensuring its dynamic stability. The disk can
approach the state of marginal stability via internal processes:
dynamic heatup, related to the development of local instabilities
both in the disk plane, and in the perpendicular direction (bending
instability)~\mbox{\cite{FKh11,Griv12}}. At the
presence of a massive spherical component (a halo, bulge), the
velocity dispersion of stars in the marginally stable disk will be
much lower than the circular velocity. However, as shown by
numerical models, it reaches  30--50\%  of circular velocity without
any external effects if the disk is self-gravitating, i.e. the mass
of spheroidal components in whose field the disk is embedded,is
relatively small (see, for
example,~\mbox{\cite{KhZT03,ZKhT04}}). Note that a
rough  mass estimate of the disks of spiral galaxies, assuming their
marginal stability at two radial scales from the center ($r\approx
2h$) is statistically consistent with the photometric estimates of
the disk mass~\cite{ZKhS11}, although for the individual
galaxies the differences may be quite large.

In general, the disk of the galaxy at any distance from the center
can have an excessive (for stability) velocity dispersion. Below we
consider the disk to be  dynamically overheated if at a sufficient
range of distances from the center  the observed velocity dispersion
of the stellar  component  $C_{\rm obs}$ is systematically higher
than the dispersion estimate  for the marginally stable disk by a
value not exceeding the characteristic scatter of points along the
radial profile of $C_{\rm obs}(r)$ (typically---10--20~km/s).
However, even in the case of a dynamical overheating of the disk, it
still makes sense to build a model of the galaxy under the
assumption of marginal stability of its stellar disk: a comparison
of this model with observations provides a constraint on the
commonly used maximum disk model, containing the ``highest
possible'' density of the disk which matches with the observed
rotation curve. This model, if it takes into account not only the
shape of the rotation curve, but also the radial profile of the
velocity dispersion of the old stellar population of the disk,
making up its bulk, may be called the refined maximum disk model.
The examples of this approach to estimate the disk mass or the disk
mass-to-luminosity ratio  are discussed, e.g., by
Bottema~\cite{Bottema88, Bottema93}, Zasov et
al.~\cite{ZKhS11,ZKhT04,Z-Sab12}, and with
regards to the LSB galaxies---by Saburova~\cite{Sab11}. If
the actual disks of galaxies have smaller masses than the mass,
ensuing from the marginally stable disk model, or if the observed
velocity dispersion of the disk stars is higher than the model
velocity dispersion projected onto the line of sight, this argues
for the dynamic overheating of the disk.

Note that the stability of the disk depends on a number of factors
difficult to account for. The analytical expressions for the
stability criterion are obtained only in the local approximation,
and only given the essentially simplifying assumptions. More
reliable results may be achieved by constructing the dynamically
stable numerical \mbox{$N$-body} models, satisfying the observed
distributions of brightness and rotation velocity, and comparing the
``model'' velocity dispersion profiles with observations. Numerical
models are particularly important for the galaxies with high
velocity dispersions of the disk stars, for which the rotation curve
of gas is absent. In this case, the observed rotation curve of stars
has to be corrected for the asymmetric drift, and this problem has
an analytical solution only in the case of a minor dispersion, when
the ratio of squares of the stellar velocity dispersion to circular
velocity is much smaller than unity.

Construction of $N$-body models for three S0 galaxies (NGC\,1167,
NGC\,4150, NGC\,6340) and an SBa-type galaxy (NGC\,2273) from the
BTA observations has been previously implemented in   Zasov et
al.~\cite{ZMKhS08}. Three-dimensional numerical dynamical
galaxy models were constructed  with the lowest possible velocity
dispersion of disk stars which provides the disk stability. The
models included a stellar disk, which is close to the maximum disk,
a bulge with an initial density distribution approximated by the
King's law, and a pseudo-isothermal halo. The spectroscopic
measurements of NGC\,6340 were later reprocessed and refined by
Chilingaryan et al.~\cite{ChNCC10}, reconfirming the
previous findings. A comparison of models with the observational
data has shown that only one galaxy out of the four modelled, the
SBa galaxy NGC\,2273 demonstrates an agreement of the observational
data with the assumption that the stellar disk is close to the
marginally stable state. In this respect this galaxy is similar to
many spiral galaxies of later types, in which the stellar velocity
dispersion is close to the stability threshold, at least at
distances of about two radial scales of the disk from the
center\cite{ZKhS11}.

\begin{figure}[h]
\setcaptionmargin{5mm} \onelinecaptionsfalse 
\centerline{
\includegraphics[width=0.33\columnwidth]{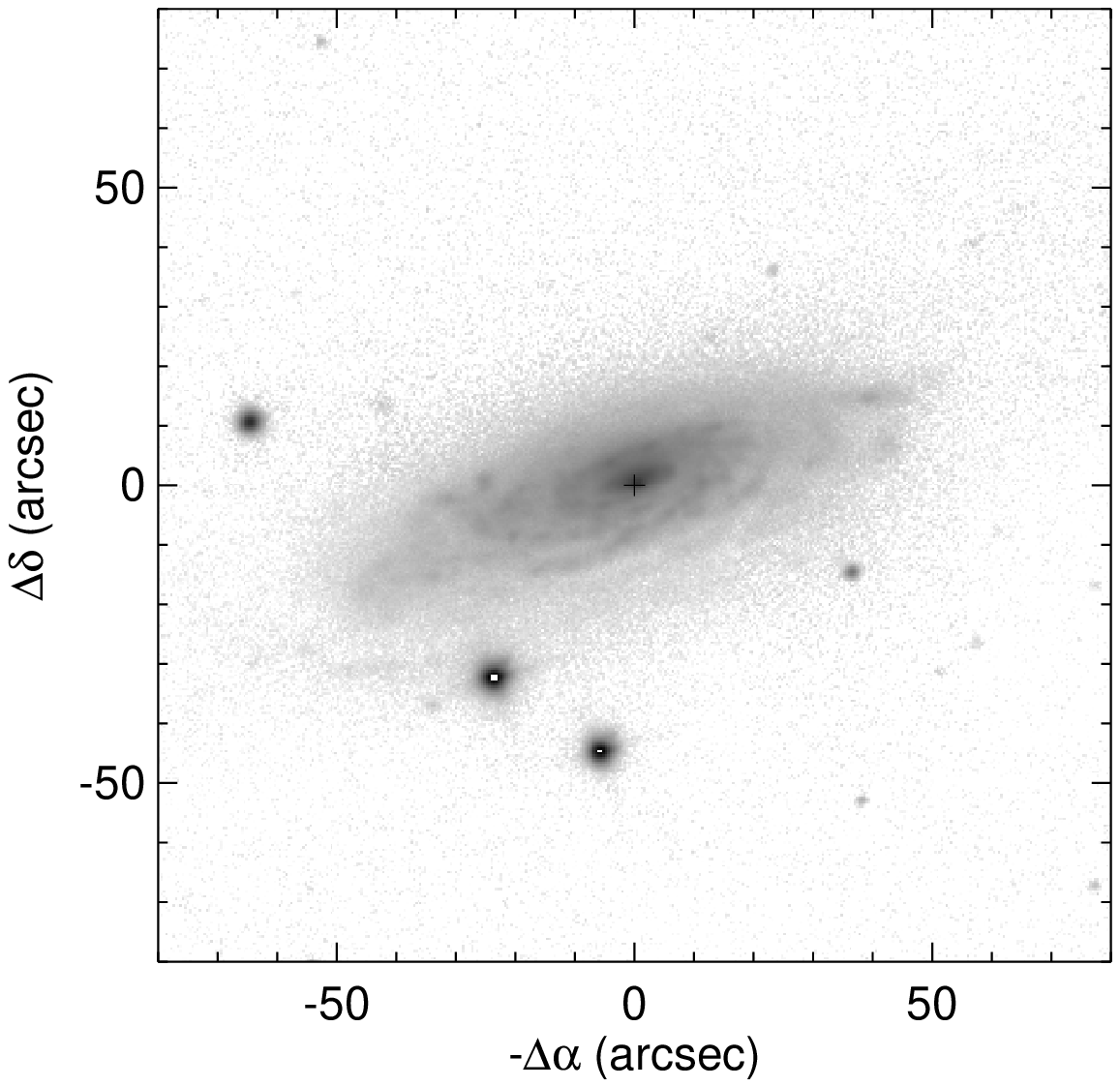}
\includegraphics[width=0.33\columnwidth]{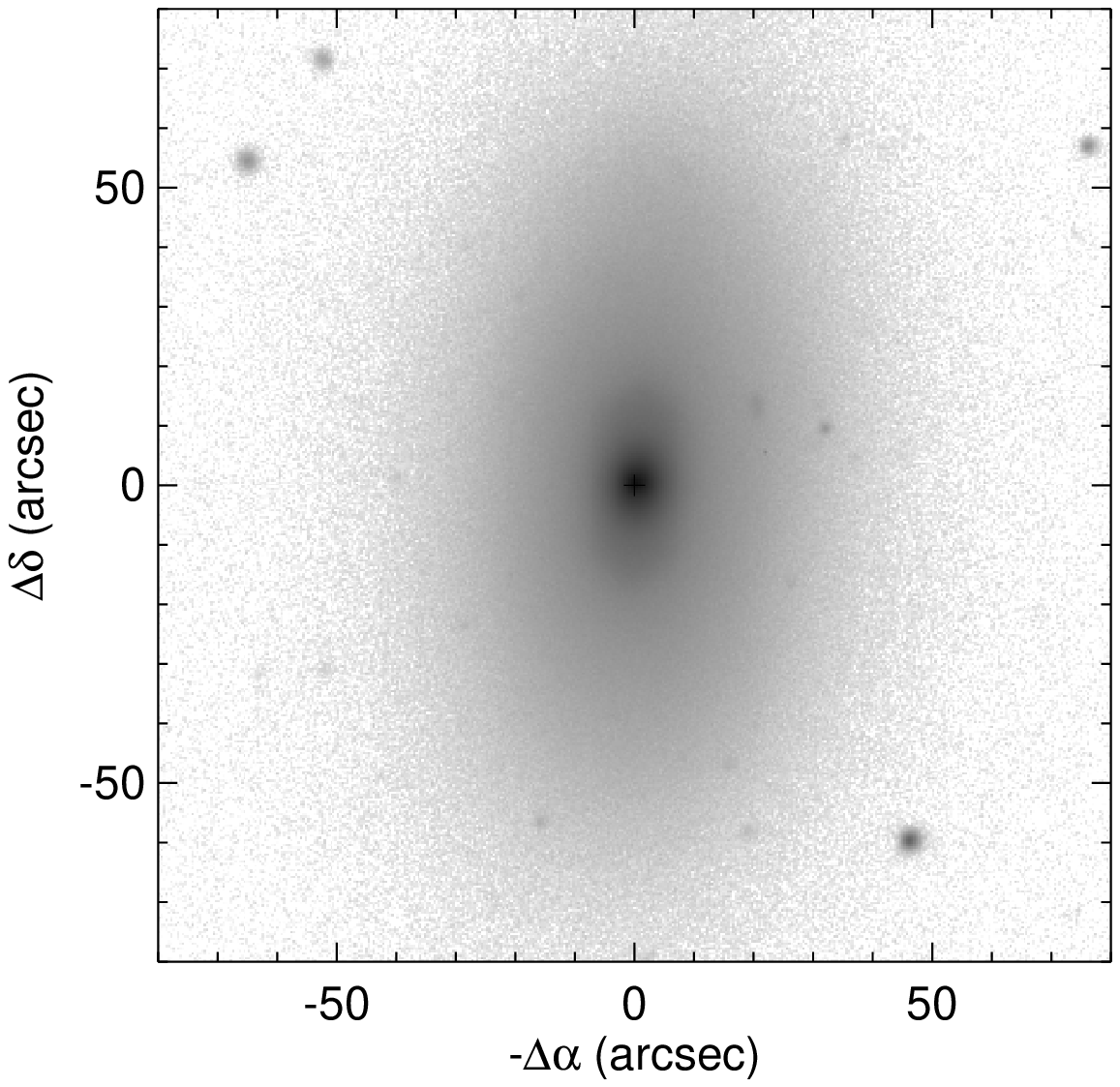}
\includegraphics[width=0.33\columnwidth]{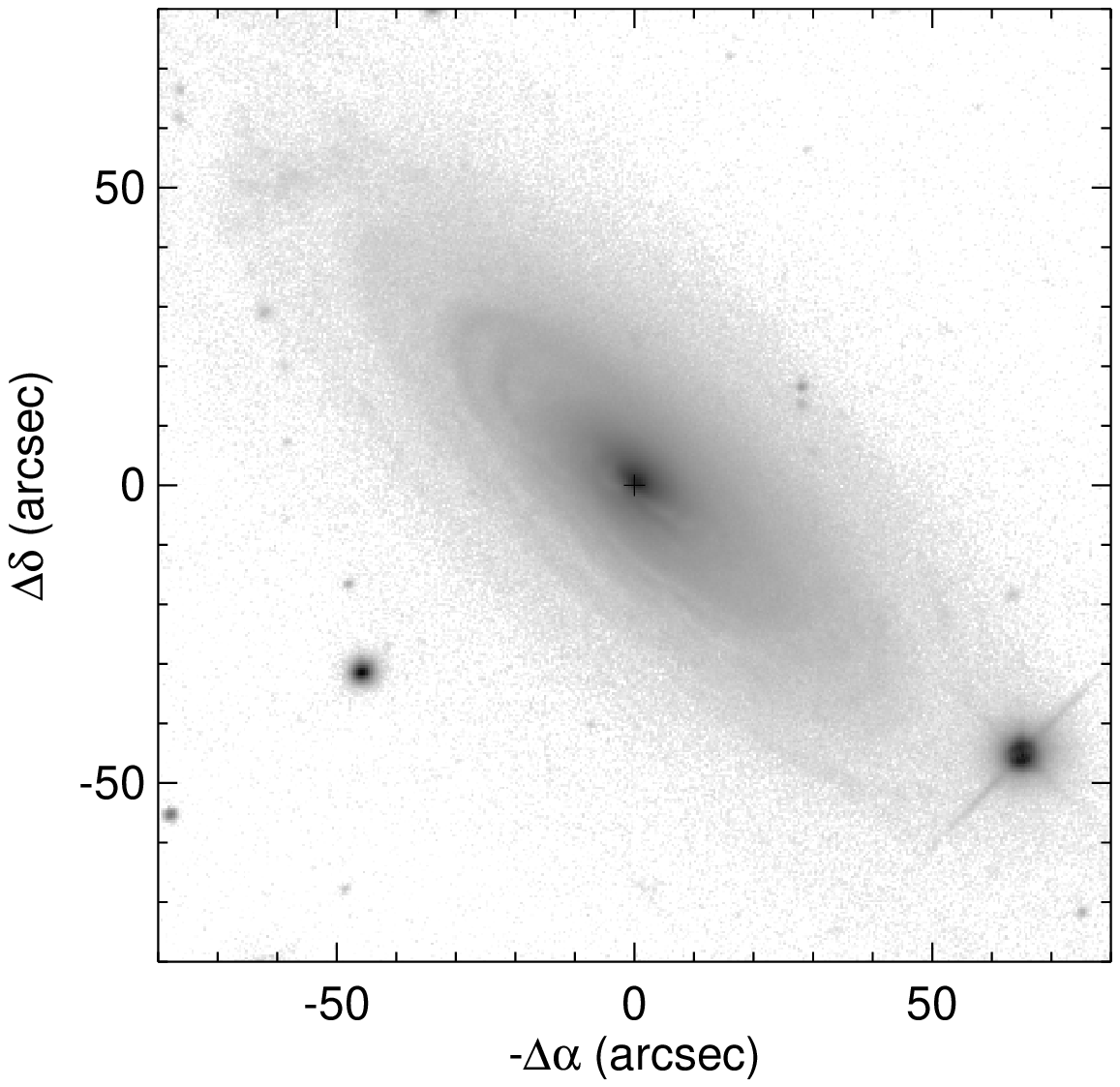}}
\captionstyle{normal} \caption{The SDSS survey images of the
NGC\,338, NGC\,3245 and NGC\,5440 galaxies in the $g$ filter.}
\label{figs_sdss}
\end{figure}

In this paper we present the results of spectroscopic observations
of three early-type galaxies: NGC\,338, NGC\,3245 and NGC\,5440
conducted at the 6-m BTA telescope of the Special Astrophysical
Observatory of the Russian Academy of Sciences (SAO RAS), and
compare the obtained velocity measurements with the numerical
models of galaxies with marginally stable disks. Their images from
the SDSS  survey are presented in Fig.~\ref{figs_sdss}.
The use of two spectral cuts made it possible to check whether we
were right in our choice of the position angle ${\rm PA}_0$ of the
dynamical major axis.

In principle,  spectral cuts along the major and minor axes of
galaxies also allow to obtain from observations the ratio of
velocity dispersions along different axes. However, for the galaxies
we consider in this paper these estimates appeared to be unreliable:
the results came out to be very sensitive to the choice of law,
approximating the observed velocity dispersion profiles along the
major and minor axes, and the rotation velocity of the stellar
component. Minor variations in the source data approximation (within
the error of observations) significantly affect the resulting
estimates of the velocity dispersion ratio $C_z /C_r$ in the
vertical and radial directions, and hence we do not give them in
this paper.

\section{DESCRIPTION OF GALAXIES}

{\it NGC\,338}. This is a type Sa or Sab galaxy with weak spiral
arms, a member of a  low-density group. In the SDSS and 2MASS survey
images the galaxy looks like an early-type disk galaxy with a bright
nucleus, extended bulge and a disk, strongly inclined to the line of
sight (Fig.~1). The SDSS images reveal in the disk an indistinctly
blurred spiral structure with separate star forming regions, notable
for their blue color. The galaxy contains a relatively large (for
lenticular galaxies) amount of H\,I ~\cite{Noord05}, so it
can be regarded as an early-type spiral galaxy with weak star
formation. The photometry and kinematics of this galaxy was
previously considered in~\mbox{\cite{Noord07,
Noord_ea07}}. The disk has a relatively low central
(deprojected) brightness and a large radial scale in the photometric
$R$-band  (about $5.8$~kpc at a distance of $65.1$~Mpc) and is
significantly inferior to the bulge in luminosity. However, as noted
by the cited authors, the conclusion about the anomalously bright
and extended bulge (the effective radius of
\mbox{$r_e\approx4.7$~kpc}) may be due to the error of the
photometric decomposition of brightness into the components within
the ``exponential disk + S\'{e}rsic bulge'' model, and to the
possible presence of dust in the center of the galaxy.

The rotation curve of the galaxy  reaches the plateau quite fast,
within a few kpc from the center~\cite{Noord_ea07}, the
maximum velocity of rotation is estimated to be  \mbox{$300 \pm
16$~km/s}~\cite{Noord_Verh07}. However, a large-scale
asymmetry of the velocity field, possibly related to the interaction
with a neighboring galaxy, has prevented the authors
of~\cite{Noord_ea07} from using the obtained velocity
estimates in the galaxy modeling. Following their work, we adopted
the position angle of major axis \mbox{${\rm PA}_0=108^\circ$} and
the angle of inclination of the disk
 \mbox{$i=64^\circ$}. Using the spectral cuts we have
constructed rotation curves both for the gas and for the stellar
disk, up to a distance of approximately $50''$ from the center.

{\it NGC\,3245}. This is a single lenticular galaxy, possessing a
bright central disk (a lens) with a sharp outer boundary at a
distance of \mbox{$r\approx15''$--$20''$} and an external main
disk. The galaxy has a very high gradient of the line-of-sight
velocity within $4''$ from the
center~\cite{Ho_Filippenko97}. There are no traces of
spiral structure, H\,II zones or star-forming regions, although we
managed to detect some faint emissions in the spectrum of the
central disk, indicating a rapid rotation within $10''$. The
galaxy has a Seyfert nucleus of a low-level activity. According to
the photometric measurements~\cite{Mendez08} the galaxy
has a compact bulge with an effective radius of \mbox{$r_e =
3\farcs3$} and a disk with a radial scale of \mbox{$h =
17\farcs6$}, or $1.5$~kpc, given the adopted distance of $18$~Mpc.
The kinematics of the galactic disk was studied
in~\cite{Simien98}, but the measurements cover only the
central $17''$, where the angular velocity is nearly constant.
From the obtained spectral cuts we have constructed the stellar
disk rotation curve, slowly receding at \mbox{$r>30''$}. We used
the photometrically determined orientation angles of the disk.

{\it NGC\,5440}. This poorly studied lenticular galaxy is a
possible member of a diffuse group. The bright inner region of the
galaxy passes into a disk of low surface brightness, in which the
low contrast arms or ring fragments are visible. The results of
photometry of the galaxy, and two-dimensional photometric
decomposition into the bulge and disk are given
in~\cite{Mendez08}. The effective radius of the bulge and
radial scale of the disk amount to $3\farcs4$ and $15\farcs7$,
respectively, which corresponds to $4.0$~kpc at a distance of
$52$~Mpc. The position angle of the photometric major axis is
\mbox{${\rm PA}_0= 46^\circ$}, which is consistent with our
measurements with two slit orientations.


\section{OBSERVATIONS AND DATA REDUCTION}

\begin{table*}
\setcaptionmargin{0mm} \onelinecaptionstrue \captionstyle{normal}
\caption{The log of observations}\label{tab1_log_obs}
\medskip
\begin{tabular}{l|lrc|c|c|c}
\hline
 \multicolumn{1}{c|}{Galaxy}   & \multicolumn{3}{c|}{Date of}        & Slit orientation & Exposure,   &  Seeing,\\
 \multicolumn{1}{c|}{name} & \multicolumn{3}{c|}{observations}  & ${\rm PA}$, deg      &  s       &   arcsec\\
\hline
    NGC\,338  & Oct &\phantom{1}2,& 2006  &  289    &    ~$7\times1200$         &   2.7\\
         & Sep &18,& 2007 &  197    &    ~$9\times1200$         &   2.4\\

    NGC\,3245 & Apr &14,& 2007 &  355    &    ~$6\times1200$         &   4.0\\
         & Apr &\phantom{1}7,& 2008  &  267    &    ~$6\times1200$         &   3.3\\

    NGC\,5440 & May &12,& 2007 &  230    &    ~$8\times1200$         &   1.3\\
         & May &16,& 2007 &  320    &    $2.7\times1200$~~   &   1.5\\
\hline
\end{tabular}
\end{table*}

The observations of galaxies were carried out in the primary focus
of the 6-m BTA telescope of the Special Astrophysical Observatory
(SAO RAS) over \mbox{2006--2007} using the SCORPIO focal
reducer~\cite{AfMois05} in the long-slit mode (with the
slit size of \mbox{$6'\times1''$}).  We used the VPHG\,2300G grism
with the operating spectral range of \mbox{4800--5550~\AA}, which
contains the absorption lines of Fe, Mg, Ti, etc., as well as the
H\,$\beta$, [O\,III], [N\,I] emissions. The   EEV\,42-40 detector
with the  CCD sized \mbox{$2{\rm K}\times2{\rm K}$} in the
\mbox{$1\times2$} binning mode   provided the $0.35''/{\rm px}$
scale along the slit. The orientation of the slit, the total
exposures and atmospheric conditions are given in the log of
observations (Table~\ref{tab1_log_obs}). The
characteristic resolution of the spectra obtained is 2.6~\AA,
which corresponds to 65~km/s in terms of velocity.

Initial data reduction included standard steps: subtracting the
averaged bias frame, accounting for the inhomogeneous illumination
and CCD sensitivity variations using the frames  of the continuous
spectrum calibration lamp (flat field),  removing the traces of
cosmic ray particles, construction of a two-dimensional dispersion
equation from the spectrum of a He-Ne-Ar calibration lamp, spectrum
linearization, summation of the spectra, subtracting the spectrum of
the night sky taking into account the instrumental profile
variations along the slit (for the details of sky subtraction
procedure, see~\cite{KatkovChil11}). The dispersion equation
was parameterized by a polynomial surface with the fifth- and
fourth-degree polynomials along and across dispersion, respectively.
The typical error of the dispersion equation is $0.03$~\AA\ for an
individual spectral line. The transition to absolute fluxes was not
performed. At each stage of the initial processing the images with
Poisson flux errors were calculated.

For the further reduction of stellar spectra the ULySS software
package~\mbox{\cite{Koleva2009ulyss,Koleva2008}}
was used, adapted to the SCORPIO data. ULySS was primarily applied
to determine the  shape of the spectrograph instrumental profile
and its variations in the frame. For this purpose we obtained the
frames of the dawn (or dusk) sky, which were  taken the same night
as the objects. To determine the variation over the field, the
spectrum of the dawn sky was split into 60 elements along the slit
and averaged for each element. The spectrum was divided into seven
overlapping segments by wavelength. After that, each part of the
spectrum was approximated by the high resolution solar spectrum
($\lambda/\Delta \lambda \approx10\,000$), convolved with the
instrumental profile in the form of the Gauss-Hermite function
with the  $v$, $\sigma$, $h3$ and $h4$
parameters~\cite{van_der_Marel1993}. The instrumental
profile was used for the  night sky spectrum  subtraction
procedure and in the analysis of the spectra of galaxies.

As the next step, the ULySS software package was used for the pixel
by pixel approximation of the observed spectra by the model spectra
of stellar populations by minimizing the  $\chi^2$ residual. The
object spectrum was adaptively binned in advance. The core of this
procedure consists in dividing the spectrum into the elements along
the slit, based on the condition that the signal-to-noise ratio in
each averaged element would be not less than the preassigned value
(typically 20--50).

Modeling the spectra,  high-resolution PEGASE.HR
models~\cite{LaBorgne2004} were used for the Simple Stellar
Population (SSP) with star formation history, described by a single
burst  and the Salpeter Initial Mass Function. Before fitting, the
PEGASE.HR model grid was subjected to the convolution with the
instrumental profile for the given slit position (for the given
bin). As the next step, within the fitting algorithm the selection
of the stellar population spectrum from the model grid was made for
the current parameters of stellar age $T$ and metallicity [Fe/H].
This was followed by the convolution with the Line-Of-Sight Velocity
Distribution function (LOSVD), which was taken in the form of a
Gauss-Hermite function. After that, the spectrum was multiplied by
the polynomial continuum, which allows to formally take into account
internal absorption in the galaxy, as well as the unaccounted
spectral sensitivity curve of the CCD detector. After fitting the
spectrum for each bin, we obtained the estimates of the
line-of-sight velocity $v_r$ (km/s), stellar velocity dispersion
$C_{\rm obs}$ (km/s), the parameters characterizing non-Gaussianity
of LOSVD: $h3$, $h4$, as well as the \mbox{SSP-equivalent} estimates
of the age $T$ (Gyr) and metallicity [Fe/H] (dex). In this paper we
used solely the estimates of the velocity and velocity dispersion. A
more detailed description of the fitting procedures and discussion
of the parameter degeneracy and algorithm stability, see the
papers~\cite{Koleva2009ulyss,CappellariEmsellem2004,Chilingarian2007}.

\section{PRINCIPLES OF MODELING}

The basic principles of constructing the numerical equilibrium
galaxy models are presented in~\cite{ZMKhS08} and in the
monograph by Fridman and Khoperskov, titled the Physics of
Galactic Disks~\cite{FKh11}. The basis of dynamic galaxy
models is the numerical integration of the equations of motion of
the  $N$-bodies, where $N$ is a few million (depending on the
model) particles, imitating the disk, bulge and the dark halo. The
solution of Poisson's equation for a given distribution of
particles was found at each integration step with the use of the
TreeCode algorithm.

We assumed that the gas rotation curve is close to the curve of
circular rotation, and the rotation velocity, determined by the
stars is lower than circular by the value of asymmetric drift. The
construction of galaxy models, in which the disks are located at the
border of gravitational stability was done via the method of
successive approximations by a gradual step-by-step increase of the
velocity dispersion of slightly unstable disk to obtain a
dynamically stable state, preserved during several revolutions of
the peripheral regions of the disk. The estimates of line-of-sight
velocities of stars and gas parallel with the stellar velocity
dispersions, obtained from the observations, were compared with
their model profiles, calculated along the major axis and projected
onto the line of sight.

To compare the observed radial profile of line-of-sight velocity
dispersion with the model profile we took into account the presence
of the bulge: a relative contribution of the disk and bulge to the
spectrum was considered to be proportional to their line-of-sight
column density at a given radial distance. The radial scale of the
disk mass distribution in all the models was assumed to be equal to
the photometrically determined brightness scale, while the
mass-to-luminosity ratio was considered as a free parameter. The
masses of the disk and other components were determined so that the
radial profile of the stellar disk line-of-sight velocity,
calculated for a given model, would be as close as possible to the
observed stellar velocity profile (i.e., not corrected for the
asymmetric drift), while the circular rotation curve would agree
with the velocity estimates, found from the emission lines, if
present. In contrast to the previous work~\cite{ZMKhS08},
the \mbox{$N$-body} model considered the bulge to be ``live'', and
its parameters were determined from the condition of the compliance
of the model velocity dispersion of the  ``disk + bulge'' system
with the observed value for the central part of the galaxy.

\section{RESULTS OF KINEMATIC ESTIMATES, COMPARISON WITH MODELS}

\begin{table}
\setcaptionmargin{0mm} \onelinecaptionstrue \captionstyle{normal}
\caption{Integral parameters of disks: results of
modeling}\label{tab2_results}
\medskip
\begin{tabular}{l|l|c|c|c}
\hline
 \multicolumn{1}{c|}{Galaxy}   & \multirow{2}{*}{Type}        & Disk mass  & \multirow{2}{*}{$M_h / M_{d+b}$}  &Disk\\
 \multicolumn{1}{c|}{name}  &   & $M_d$, $10^{10}~M_\odot$  &  &    state\\
\hline
    NGC 1167 & S0  & 39    & 0.51     & o\\
    NGC 2273 & SBa & 8.7  & 0.65       & m\\
    NGC 4150 & S0  & 0.53 & 0.56  &   o\\
    NGC 6340 & S0  & 4.54 & 0.66   & o\\
\hline
    NGC 338   & Sa  & 15.4 & 0.32     & m\\
    NGC 3245 & S0  & 1.1:   & 0.6:     & m\\
    NGC 5440 & Sa  & 19.3 & 0.56  &  o\\

\hline
\end{tabular}
\end{table}

\begin{figure}[h]
\setcaptionmargin{5mm} \onelinecaptionsfalse \centerline{
\includegraphics[width=0.5\textwidth]{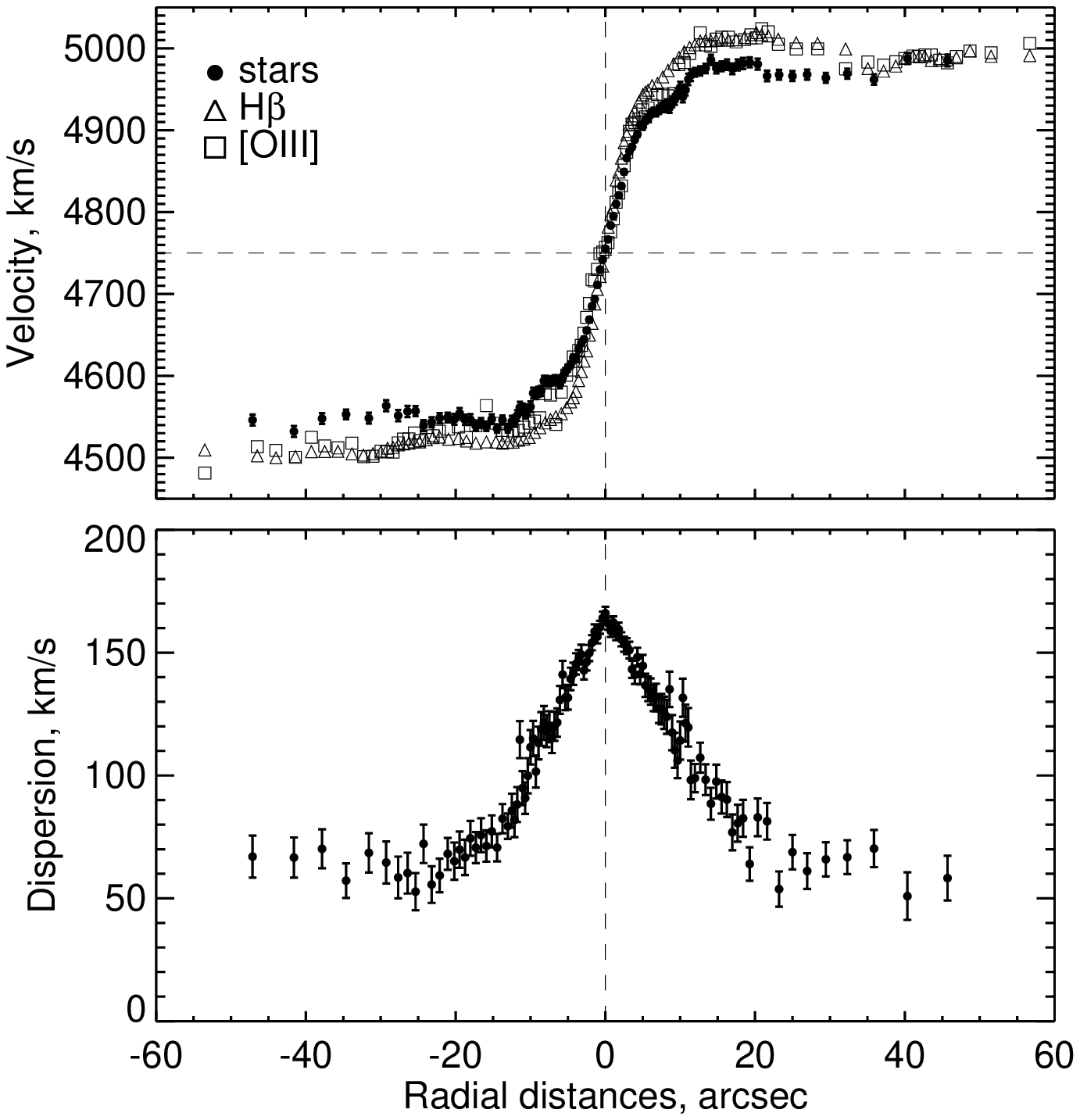}
\includegraphics[width=0.5\textwidth]{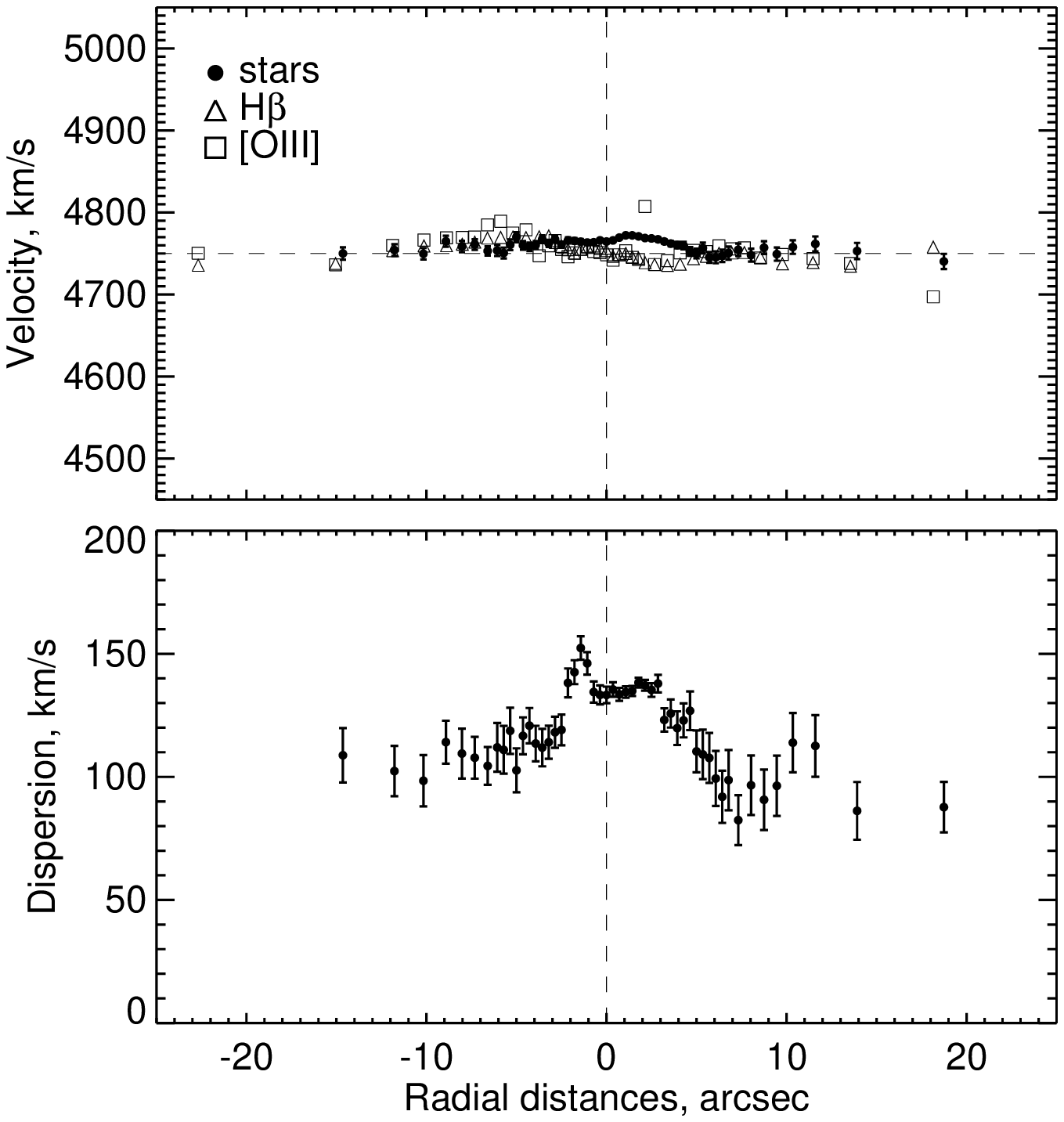}
} \captionstyle{normal} \caption{The  NGC\,338 galaxy. Left:
results of measurements along the major axis, right: measurements
along the minor axis. Top: radial profiles of  line-of-sight
velocities of stars and ionized gas, bottom: stellar velocity
dispersion. The error bars of line-of-sight velocities of ionized
gas are not shown because their size is not larger than the size
of icons.} \label{figs_radial_profiles_n338}
\end{figure}

\begin{figure}[h]
\vspace{3mm} \setcaptionmargin{5mm} \onelinecaptionstrue
\centerline{
\includegraphics[width=0.5\textwidth]{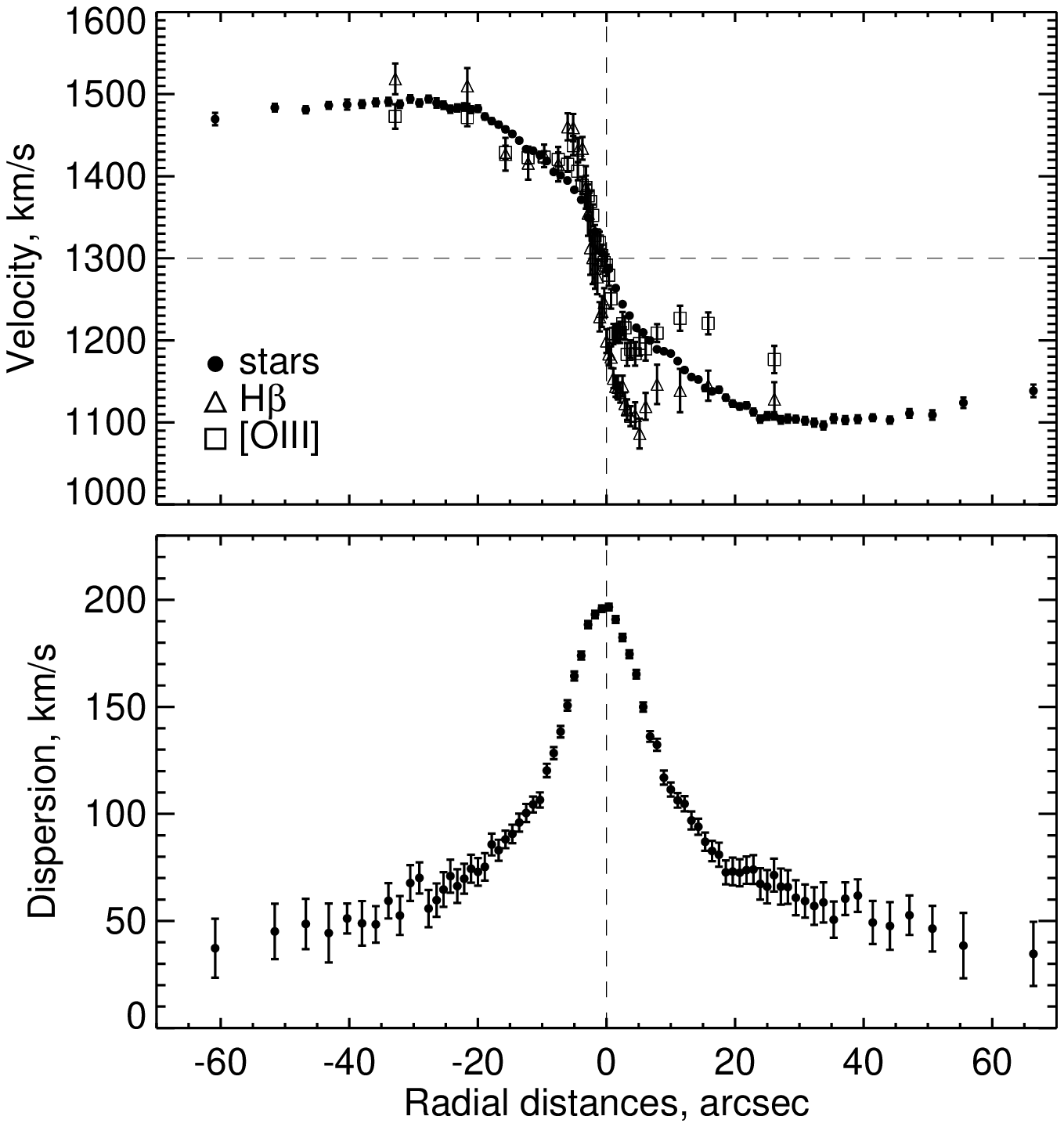}
\includegraphics[width=0.5\textwidth]{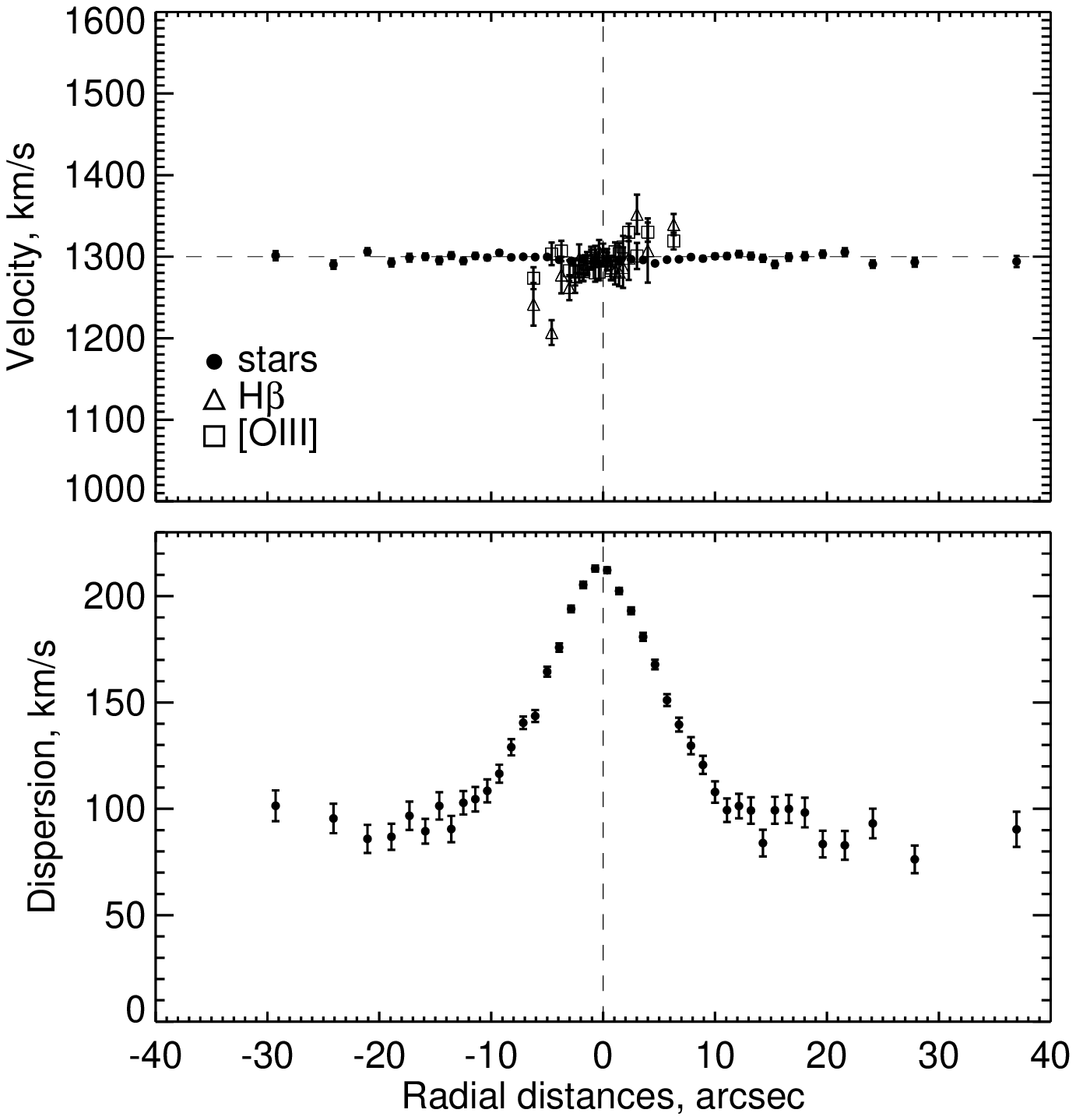}
} \captionstyle{normal} \caption{The same as in
Fig.~\ref{figs_radial_profiles_n338}, but for the
NGC\,3245 galaxy.} \label{figs_radial_profiles_n3245}
\end{figure}

\begin{figure}[h]
\setcaptionmargin{5mm} \onelinecaptionstrue \centerline{
\includegraphics[width=0.5\textwidth]{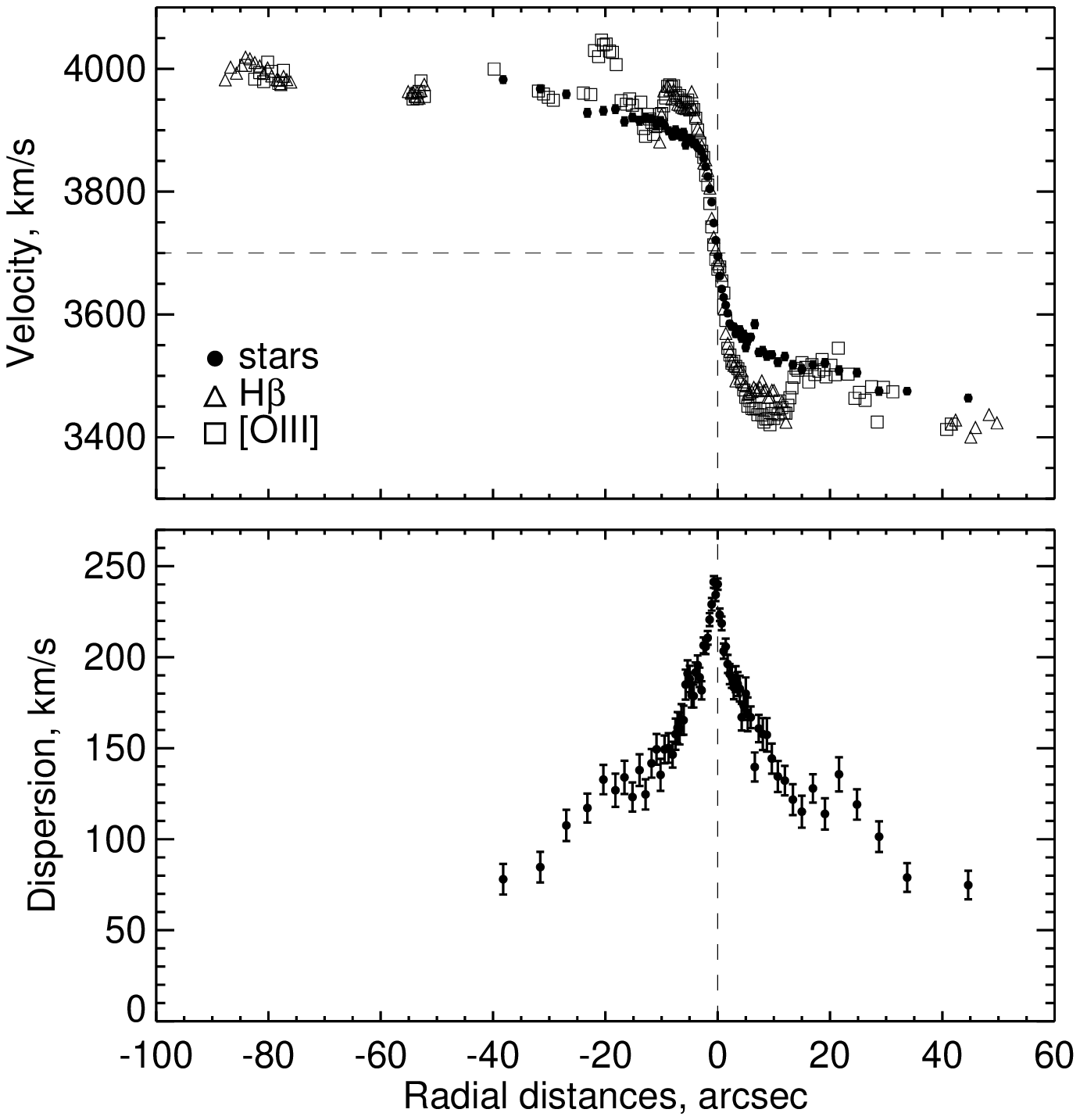}
\includegraphics[width=0.5\textwidth]{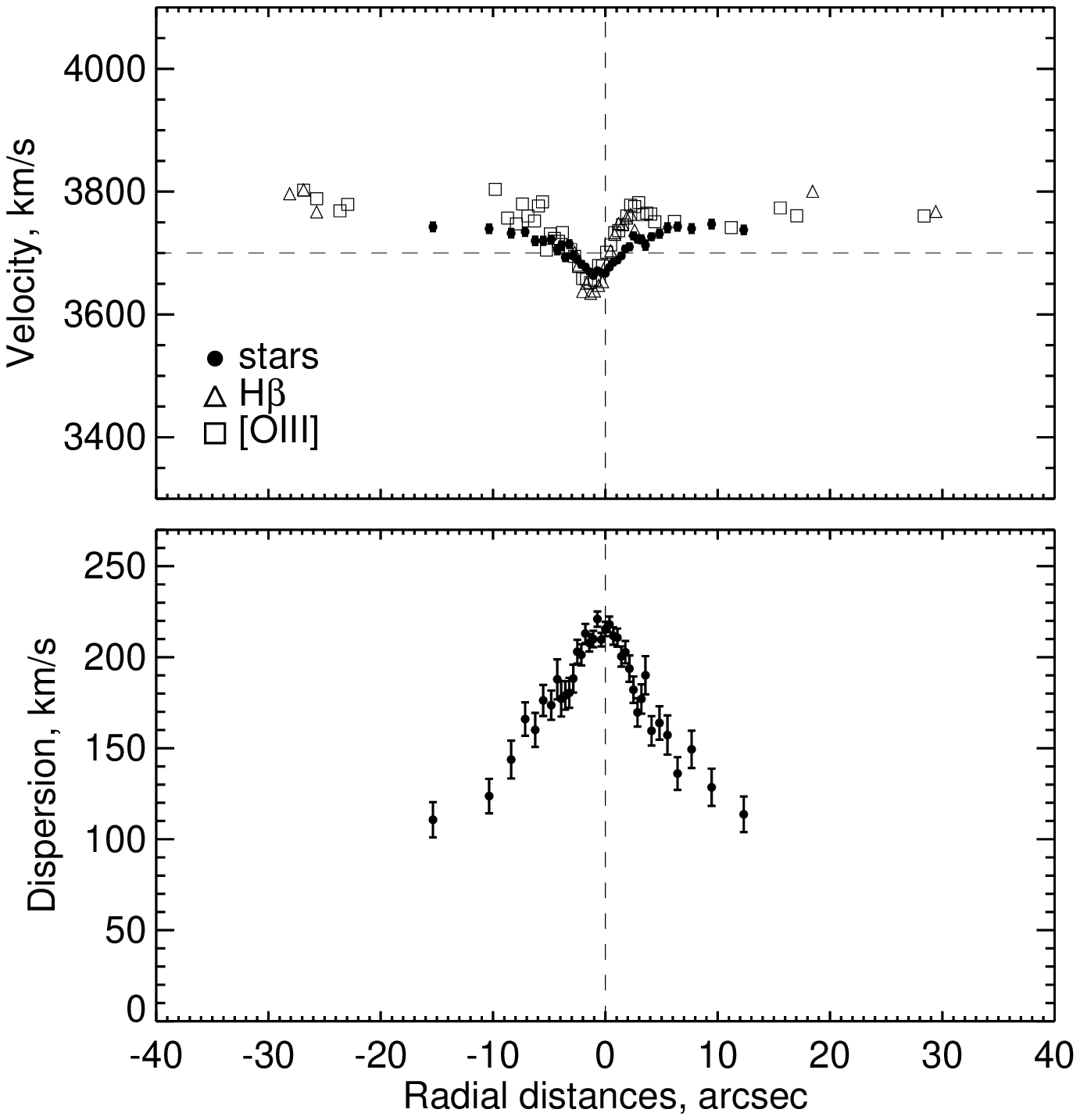}
} \captionstyle{normal} \caption{The same as in
Fig.~\ref{figs_radial_profiles_n338}, but for the
NGC\,5440 galaxy.} \label{figs_radial_profiles_n5440}
\end{figure}

\begin{figure}[h]
\setcaptionmargin{5mm} \onelinecaptionsfalse \centerline{
\includegraphics[width=0.33\textwidth]{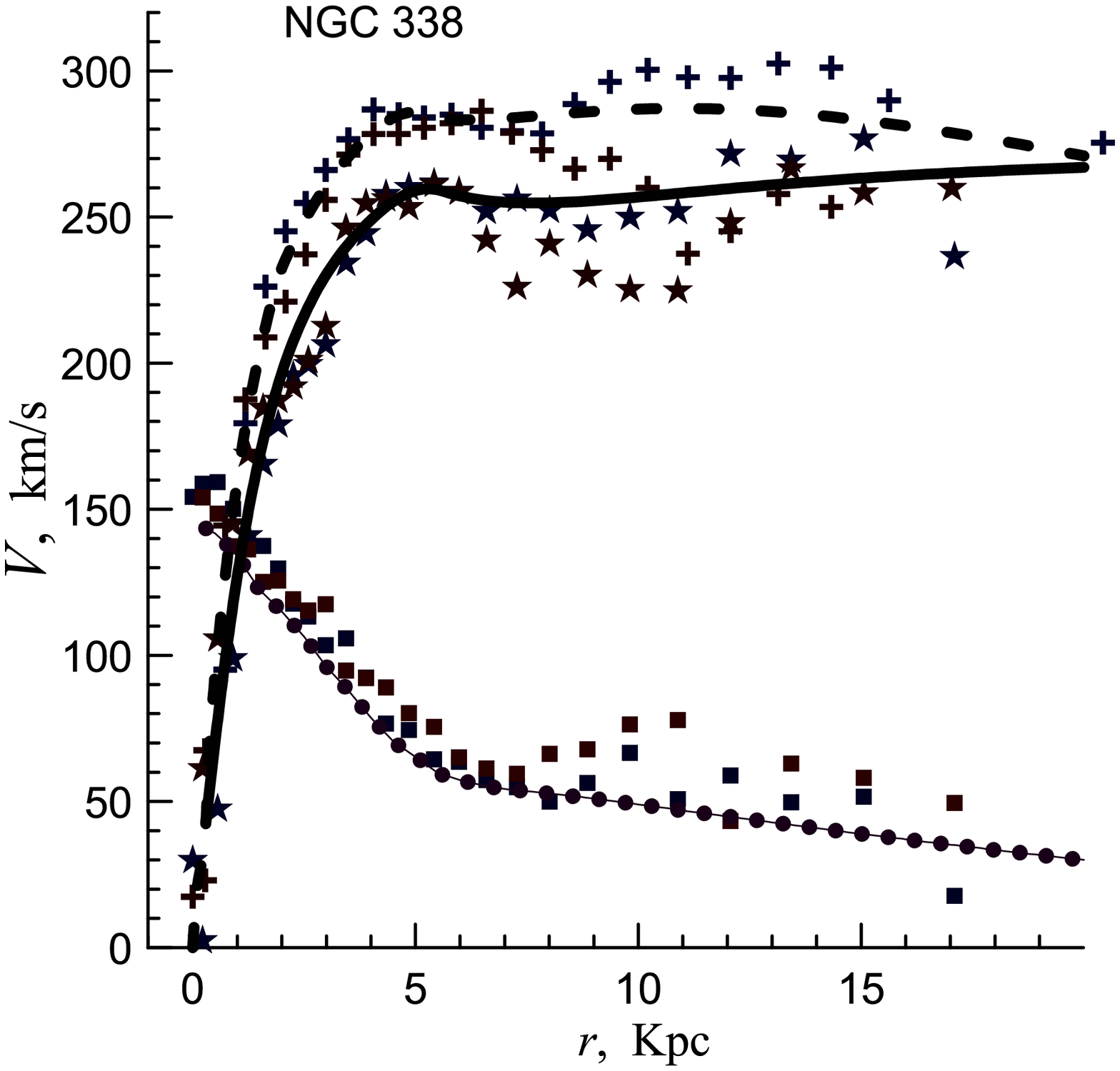}
\includegraphics[width=0.33\textwidth]{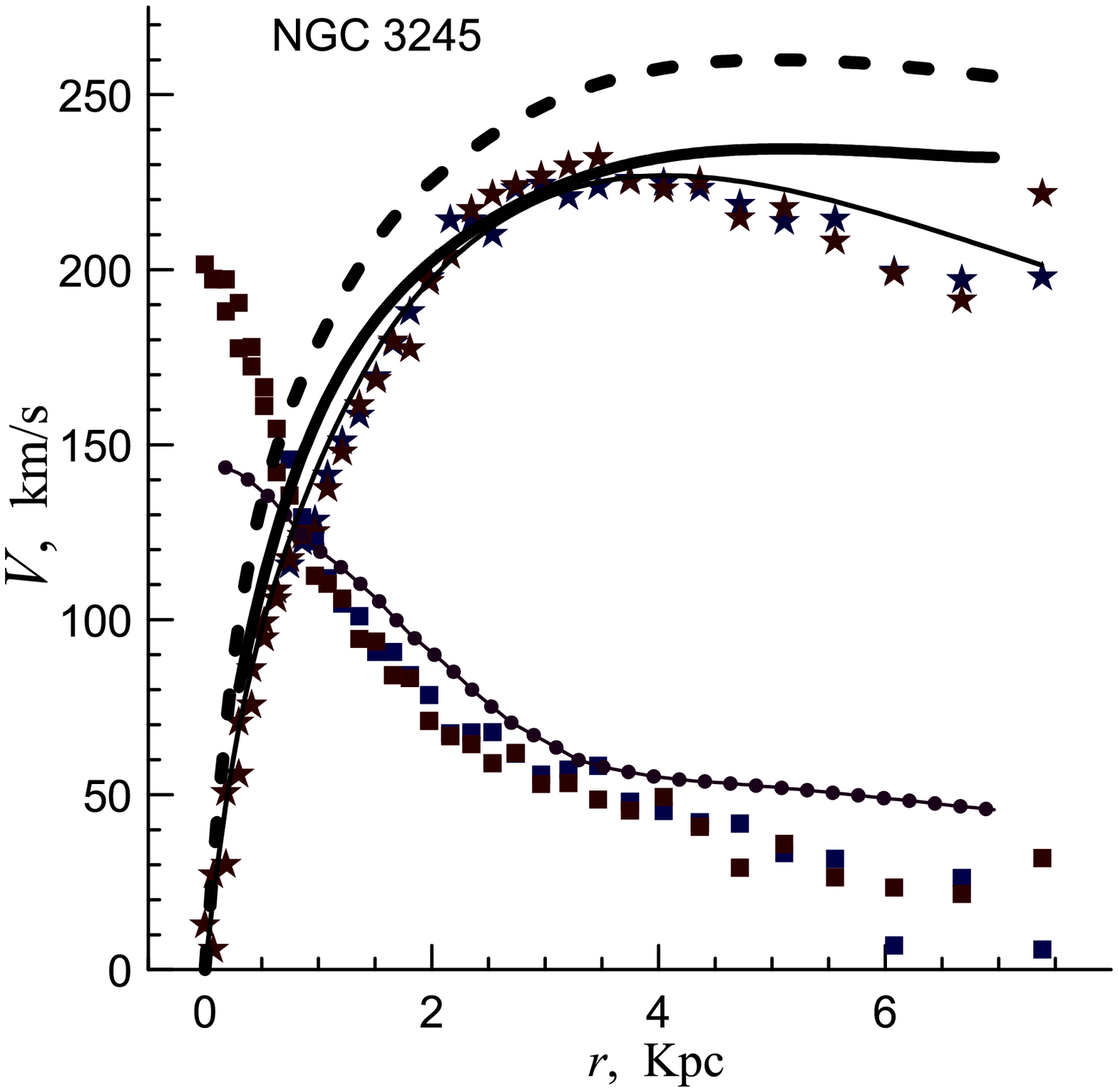}
\includegraphics[width=0.33\textwidth]{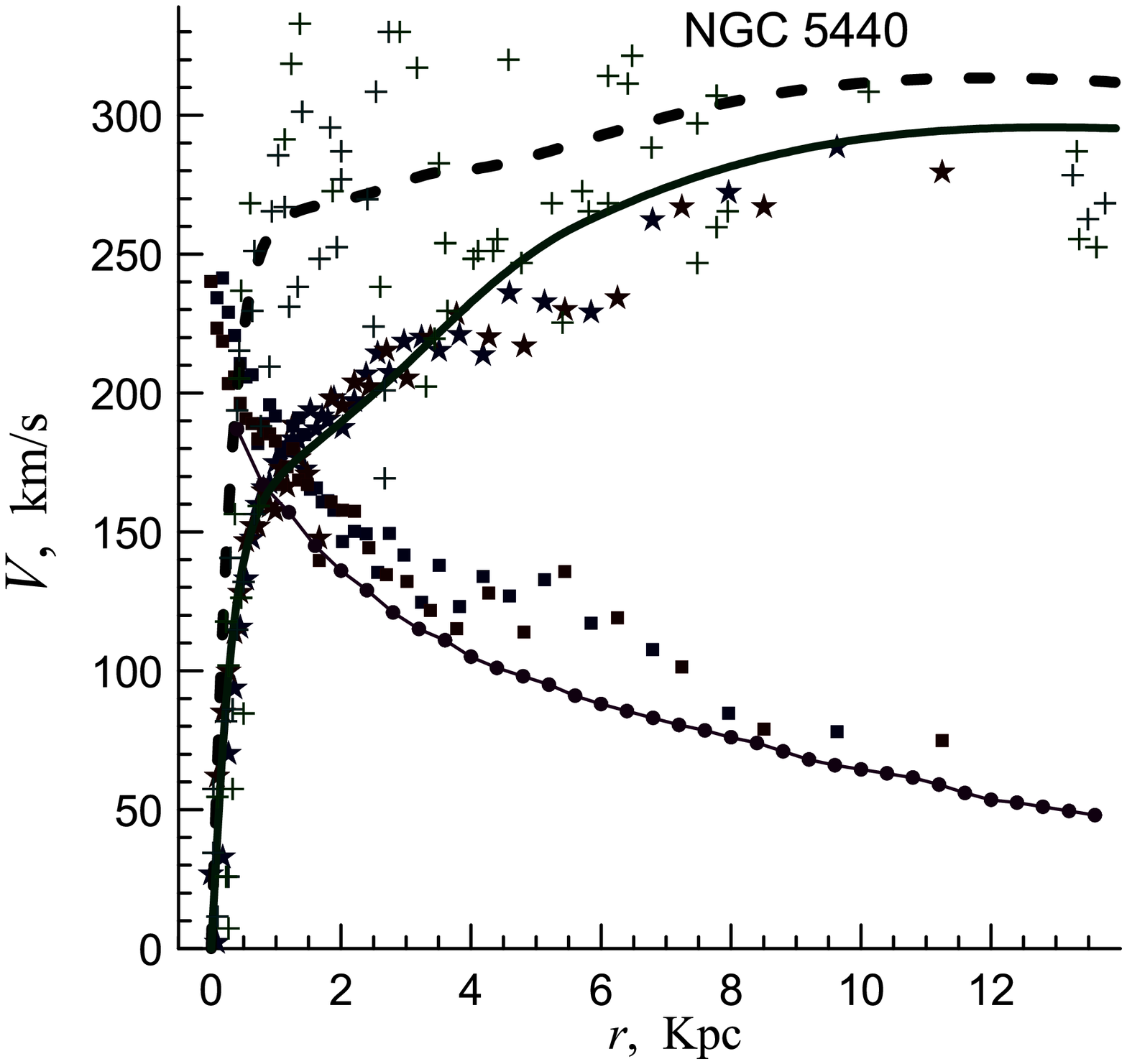}}
\captionstyle{normal} \caption{The results of   dynamic modeling.
The thick dashed line describes the circular rotation velocity,
the thick solid line---stellar rotation velocity  in the model,
the stars---stellar rotation velocity from observations,
crosses---gas rotation velocity from observations,
squares---stellar velocity dispersions from observations, the thin
line with small circles---velocity dispersion along the line of
sight in the marginally stable disk model. For NGC\,3245 two model
stellar rotation curves are given (the thin and thick lines),
corresponding to different velocity dispersions along the
$z$-coordinate (see text). } \label{figs_models}
\end{figure}

The obtained radial profiles of the line-of-sight velocities and
stellar velocity dispersions    are demonstrated in Figs.~2--4. The
reconstructed rotation curve for a marginally stable disk of each
galaxy (i.e. for a stable disk model with the maximal density
possible), as well as the model and observed radial profiles of
stellar velocity dispersions are shown in
Fig.~\ref{figs_models}. The resulting estimates of the
parameters of galaxy components are given in
Table~\ref{tab2_results}. The photometric parameters of
galaxies, if their source is not indicated, are taken from the NED
database\footnote{\tt http://ned.ipac.caltech.edu/}. The
three-dimensional axisymmetric model includes a truncated King's
bulge with a fixed total mass $M_b$, an exponential disk and a
spherical pseudo-isothermal halo. The solid and dashed lines in
Fig.~\ref{figs_models} describe the model curves of circular
velocity and stellar disk velocity, respectively. The descending
solid line in the bottom part of the figure demonstrates the radial
profile of the stellar disk velocity dispersion projected onto the
line of sight for the model constructed. Although all the three
galaxies belong to the early morphological types, their rotation
curves turned out to be different in shape, indicating different
mass distributions in galaxies. The NGC\,338 galaxy reveals the
rotation curve which comes to a plateau at $r\approx 3$~kpc, and
displays noncircular velocities at large $r$. In NGC\,3245 the
rotation curve for the stars passes through a maximum, while in
NGC\,5440 the curve is notable for a very high velocity gradient in
the central region, which gradually decreases with distance from the
center.

The ``live'' bulge was built from dynamic particles with the same
masses as for the disk. In an iterative process of the construction
of an equilibrium marginally stable disk, the parameters, describing
the density distribution in the bulge remained the same, but the
bulge velocity dispersion slightly varied and its flattening
slightly increased due to the gravitational effects that the disk
imposes on the bulge.  During the galaxy modeling, a slight
redistribution of matter took place in the bulge, especially in the
centermost region, where the density increased.  However these
changes did not have a significant impact on the dynamics of the
disk. The parameters of the bulges of galaxies from the models
constructed (the mass, scale, and the radius of the bulge in King's
model) are listed in Table~\ref{tablbulge}.

\begin{table}
\setcaptionmargin{0mm} \onelinecaptionstrue \captionstyle{normal}
\caption{Bulge parameters}\label{tablbulge}
\medskip
\begin{tabular}{l|c|c|c}
\hline
\multicolumn{1}{c|}{Galaxy}   & Bulge mass      & Scale     & Radius  \\
\multicolumn{1}{c|}{name} &  $M_b$, $10^{10}~M_\odot$   & $r_b$, kpc  & $r_b^{\max}$, kpc\\
\hline
    NGC\,338   & 4.80  & 1.12 & 6   \\
    NGC\,3245  & 0.45 & 0.35 & 1.6 \\
    NGC\,5440  & 3.20  & 0.38 & 3.6 \\

\hline
\end{tabular}
\end{table}

Let us give comments on each galaxy.

{\it NGC\,338}. This is a fast rotating galaxy, its circular
velocity reaches \mbox{280--300~km/s}. The profile of line-of-sight
velocities was obtained both from the absorption lines, and
emissions in H\,$\beta$, [N\,I] and [O\,III]
(Fig.~\ref{figs_radial_profiles_n338}). The observations of
the galaxy were carried out with the slit orientations $17^\circ$
and $108^\circ$ (the latter angle is close to the major axis
position angle). The rotation curve reaches saturation at
\mbox{$r\approx 4$~kpc} ($12''$), however at \mbox{$r > 8$~kpc}
($25''$) rotation velocities at the opposite sides from the center
differ. This is especially noticeable in the curves
along the major axis, where the velocity difference (taking into
account the projection effects) reaches~50 km/s. This confirms the
asymmetry of the velocity field,  earlier found by Noordermeer and
van~der~Hulst~\cite{Noord_ea07}, being the most prominent in
the H\,I line. Our measurements show that this anomaly affects not
only the gas, but the stellar component also. At the same time, the
decrease in the rotation velocity of stars is compensated by
increased velocity dispersion, indicating that this anomaly does
really exist. One may propose that its reason is due to the
perturbation of the velocity field of stars and gas in the process
of merger of a small satellite galaxy. However without the data on
the velocity field of the stellar disk we can not perceive whether
this effect is local or the anomaly covers a larger area of the
disk.

The velocity of rotation of the gas, determined from the emission
lines, markedly exceeds the velocity of rotation of the stellar
disk, which may be explained, both qualitatively and quantitatively,
by a higher stellar velocity dispersion. The observed velocity
dispersion, which is particularly high within the \mbox{$r\leq
15''\approx 5$~kpc} region, dominated by the galactic bulge, rapidly
decreases and reaches a plateau at \mbox{$r\approx 20''$}. Along the
minor axis \mbox{(${\rm PA}=17^\circ$)} the velocity dispersion
profile can be traced more poorly. The dispersion remains high up to
\mbox{$r \approx 10''$--$15''$}, being determined mainly by the
contribution of the bulge.

A comparison of the stellar line-of-sight velocity dispersion
profile, obtained from observations  (squares)  with that calculated
for the marginally stable disk  model (the bottom curve in
Fig.~\ref{figs_models}, left) shows that the observed
velocity dispersion varies along the radius, following the model
profile, and only slightly (on the average by 10--20~km/s) exceeds
the model estimates. The typical dispersion errors have roughly the
same value. We can therefore conclude that the disk within the
radius of \mbox{$7$--$8$~kpc} \mbox{($22''$--$25''$)} is close to
the marginally stable state, and hence was not subjected to a strong
dynamic heating. At larger distances, the majority of velocity
dispersion estimates lie higher than the model dependency, which
suggests a noticeable overheating of the disk.

{\it NGC\,3245}. The galaxy turned out to be very difficult to
simulate. Moreover, we were unable to develop an adequate
self-consistent model that would properly describe the kinematics of
the entire disk: the external and internal regions of the disk
deviate from the model of the maximal marginally stable disk, though
for different reasons, and therefore require separate discussion.
Let us consider this galaxy in more details.

The profiles of the rotation velocity and velocity dispersion of the
galaxy have a symmetrical form. However, the galaxy has a number of
dynamic features. Firstly, within the radius \mbox{$10''$--$15''$}
\mbox{($\approx1$~kpc)} from the center the
 rotation velocity of the gas (projected onto the line of sight)
exceeds the velocity of the stellar disk. It allows to assume that
the gas forms a dynamically isolated disk, inclined with respect to
the stellar disk, since a cut along the minor axis of the galaxy
exhibits a non-zero gradient of the gas velocity
(Fig.~\ref{figs_radial_profiles_n3245}). It is impossible to
get a reliable estimate of this gradient from the data available,
since the emissions are observed only near the nucleus.

Another feature of this galaxy is a peak revealed on the stellar
disk rotation curve: starting at \mbox{$r\approx4$~kpc} (about
$45''$) the velocity of its rotation is declining. The velocity
dispersion ratio \mbox{$C_z / C_r \approx 0.5$--$0.7$},  typical of
galactic disks (in agreement with numerical models which are stable
with respect to the bending perturbations) does not allow to
reproduce the observed peak of rotation. To explain the shape of
this curve, we have to assume that the region of the peak and the
subsequent rotation velocity decline satisfies the condition
\mbox{$C_z \approx C_r$}, i.e., the disk has a relatively large
vertical velocity dispersion,  which makes the thickness of the
stellar disk to be very large (a scaleheight is about $1.2$~kpc near
the peak of the curve, which approximately corresponds to the height
of the thick disk of our Galaxy). Moreover,  the vertical scale of
the stellar disk should further increase with radius as a result of
the surface density decrease in the disk. The stellar disk rotation
curve, corresponding to this model, is shown by a continuous thin
line in Fig.~\ref{figs_models}(middle). However, the
circular velocity curve of the galaxy (dashed line in
Fig.~\ref{figs_models}) may have no peak, and apparently
reaches a plateau.

\begin{figure}[h]
\setcaptionmargin{5mm} \onelinecaptionsfalse
\includegraphics[width=0.75\columnwidth]{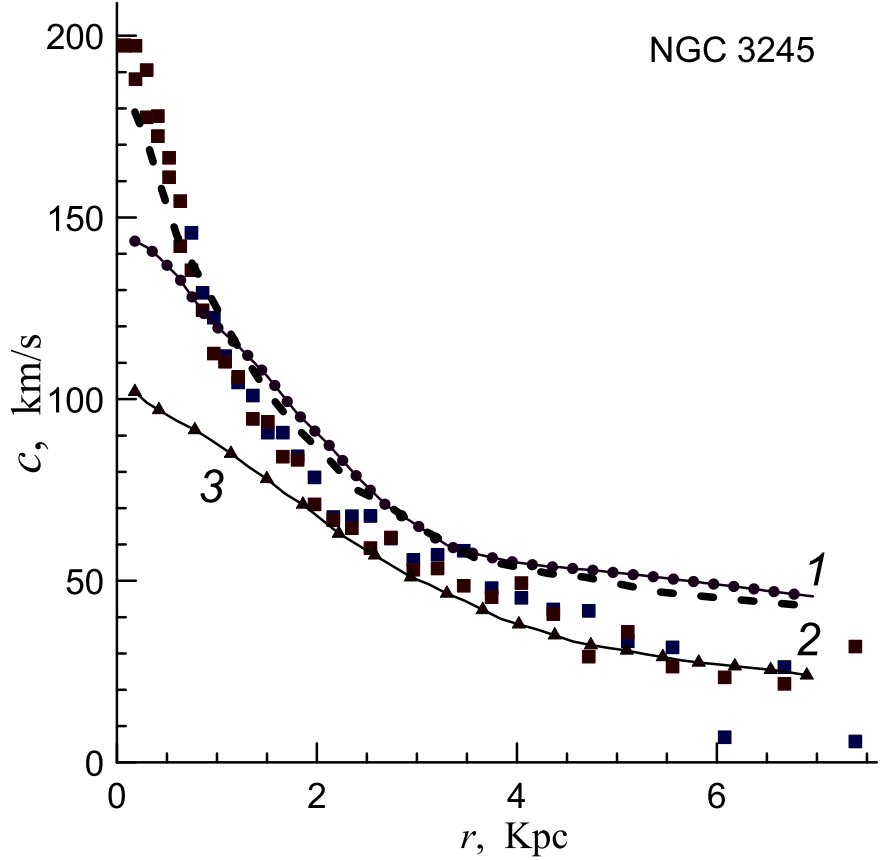}
\captionstyle{normal}
\caption{The radial distributions of the velocity dispersion along
the line of sight for the  NGC\,3245 galaxy. The observations are
marked by squares, curve (\textit{1}) corresponds to the model
shown in Fig.~5, the dashed line (\textit{2})---model with an
overheated central part of the disk, (\textit{3})---model with a
massive halo and a lightweight, marginally stable disk.}
\label{figs_N3245}
\end{figure}

The third feature of the galaxy is high velocity dispersion  of
stars in the central region, not simulated within a simple model
with the lowest possible velocity dispersion. The observed velocity
dispersion of disk stars begins to increase towards the
center---even before reaching the area, dominated by the bulge. In
Fig.~\ref{figs_N3245}, in addition to the model of maximal
marginally stable disk (curve~(\textit{1})), a dashed line
(\textit{2}) shows the results of the model, where the disk density
is lower (that is the radial velocity dispersion exceeds the minimum
value, required for disk stability), but the central part of the
disk is overheated. This inner extended region of the disk with high
stellar velocity dispersion can be connected with the central
thickening of the disk (a lens, a pseudo-bulge) of this galaxy.

Note, that out of all the galaxies we examined, NGC\,3245
is the only one for which the refined maximal disk model predicts
over a large distance interval a higher than observed stellar
velocity dispersion: by \mbox{10--20~km/s} for \mbox{$r=
1$--$4$~kpc} (around \mbox{$10''$--$45''$}) and by
\mbox{25--30~km/s} for \mbox{$r=6$~kpc} (around $70''$) (see
Fig.~\ref{figs_models}). Since the velocity dispersion of a
marginally stable disk is minimally acceptable, we must conclude
that in this case, the model of a maximal disk overestimates its
density. Outside \mbox{$r=2$~kpc} (around $20''$)  the best
agreement between the observed velocity dispersion with that
expected for a marginally stable disk (curve~(\textit{3}) in
Fig.~\ref{figs_N3245}) is achieved at the total mass of the
disk of \mbox{$M_d=1.1\times 10^{10}~M_\odot$}, what is roughly
twice smaller than for the maximal disk.

Thus, the dynamic features of the observed galaxy can be explained
if we accept that its disk is ``overheated''  in the central part,
whereas at larger distances it may be marginally stable to the
perturbations in the disk plane, having at the same time a high
\mbox{$C_z/C_r$} ratio, and hence, a large thickness.

{\it NGC\,5440}. The line-of-sight velocity curves for this galaxy
are obtained both for the stars and gas from the [N\,I] and [O\,III]
lines (Fig.~\ref{figs_radial_profiles_n5440}). The velocity
distribution along the major axis demonstrates a regular nature of
the disk rotation, especially for the stellar component, where
within the growth  segment of the rotation curve the stellar
component  rotates slower, and velocity varies monotonically, while
the velocity of the gas component passes through a peak at the
distance of about $10''$ from the center. In the plateau segment
\mbox{($r > 20''$)} the velocities of stars and gas even out. The
central minimum in the distribution of velocities of gas and stars
along the minor axis is an artifact, caused by the slight shift of
the slit, which did not pass exactly through the center of rotation.
At the presence of a steep velocity gradient, the offset can be as
little as $1''$--$2''$ to produce it.

The model velocity distributions are compared with the observations
in Fig.~\ref{figs_models} (right). The rotation curve is
determined much more confidently from the stars than from gas: the
spread of the gas velocity estimates reaches 50~km/s relative to the
rotation curve, only slightly decreasing towards the periphery.
Surprisingly, noncircular motions of gas is observed in spite of the
low intensity of star formation, which could have sustained them.

To build a dynamic model of the galaxy, it is preferable to use the
velocity estimates of the stellar component, the radial profile of
which has a smooth and symmetrical form. As can be seen from the
figure, line-of-sight velocity dispersion of stars exceeds the
expected value for the maximal marginally stable disk model by
20--40~km/s, and this takes place over the entire obtained rotation
curve. Therefore, the stellar disk of the galaxy can be considered
as moderately overheated.

\section{DISCUSSION AND CONCLUSIONS}

Out of seven early-type disk galaxies considered in this paper and
in~\cite{ZMKhS08}, the model of a marginally stable disk
can explain the observed velocities and velocity dispersions in
only two galaxies: NGC\,2273 and NGC\,3245, and in the inner
region of NGC\,338 (see Table~\ref{tab2_results}, where
the last column contains the indications of an overheated disk
(``o'' for overheated), or a disk for which the observations are
consistent with the hypothesis of its marginally stable state
(``m'' for marginal)). Both NGC\,2273 and NGC\,3245 are isolated
lenticular galaxies, and the probability of dynamic heating from
the environment is minimal, while the remaining galaxies belong to
rarefied groups. As for NGC\,338, it is not a ``classical''
lenticular galaxy, since the spiral disk structure is observed
there and it is likely to be an early type spiral.
Table~\ref{tab2_results} also lists the mass of the disk
and the relative halo mass within four radial scales of the disk.
The range of values of the latter is quite typical for disk
galaxies.

The results of observations and numerical models of the S0--Sa
galaxies confirm the conclusion that a significant proportion of
the early-type disk galaxies (but far not all of them) have
dynamically overheated stellar disks at the distance of 1--2
radial scales from the center, meaning that their velocity
dispersion is significantly higher than that required for the
gravitational stability of the disk. Note that the galaxies with
overheated disks do not excel neither by the disk mass, nor by
the relative mass of the halo, and nor by disk mass-to-luminosity
ratio (see Table~\ref{tab2_results}), which argues for
the external mechanisms causing the velocity dispersion to
increase. Apparently, these galaxies have undergone in the past a
stage of mergers with smaller systems, which led to a rapid
reduction of the amount of gas in the disk through  gas depletion
and/or a stimulation of an outburst of activity in the nucleus,
able to sweep the gas out of disk. As a number of spectroscopic
observations of lenticular galaxies show, their stellar disks are
characterized by a high magnesium-to-iron ratio, which makes them
similar to the thick disks of spiral
galaxies~\cite{sil12a, sil12b}. Then, dynamic
heating of the stellar disk would have to take place in the first
1--2~billion years after the trigger of the violent stage of its
formation, that is, before the interstellar gas got enriched with
iron as a result of activity of the Type I supernovae. In
contrast to lenticular galaxies, the galaxies of later
morphological types rarely have an excess of stellar velocity
dispersion at a distance of several radial scales from the
center~\cite{ZKhS11}. Therefore, the early evolution of
spiral galaxies generally advanced less vigorously.

A part of lenticular galaxies have avoided intense dynamical heating
of the disk. In these cases, the morphological features inherent to
this type of  galaxies appear to be related to other events, such as
a high activity of the nucleus, which may affect the abundance of
gas and star formation in the galaxy~\cite{Silk09}, or to
the interaction with intergalactic gas, if the galaxy is located in
a cluster or a compact group. It is noteworthy that two of the most
isolated galaxies contained in our small sample (NGC\,2273 and
NGC\,3245), are precisely the objects where we have reasons to
suppose the presence of marginally stable disks. We can therefore
conclude that there are at least two different mechanisms,
responsible for the formation of lenticular galaxies, and the major
part of these galaxies appears to have undergone an early stage of
formation of a dynamically ``hot'' disk.

\begin{acknowledgments}

The work was supported by the Ministry of Education and Science of
the Russian Federation. The results of observations were obtained
with the 6-m BTA telescope of the Special Astrophysical
Observatory Academy of Sciences, operating with the financial
support of the Ministry of Education and Science of Russian
Federation (state contracts no.~16.552.11.7028, 16.518.11.7073).
The authors are grateful to A.~V.~Moiseev for the help in
conducting observations and valuable suggestions, as well as the
referee, who made a number of useful comments. The authors also
express appreciation to the Large Telescope Program Committee of
the RAS for the possibility of implementing the program of
spectroscopic observations at the BTA. We as well made use of the
SDSS, supported by the Alfred~P.~Sloan Foundation, the
participating institutions, the National Science Foundation, and
the U.S. Department of Energy. This work was supported by the
Russian Foundation for Basic Research grants (project
nos.~11-02-12247-ofi-m, 12-02-00685), the Presidential Grant
\mbox{no.~MD-3288.2012.2}, and Dmitry Zimin's non-profit Dynasty
Foundation. The authors are grateful to SRCC of the Moscow State
University for the use of Chebyshev and Lomonosov supercomputers.

\end{acknowledgments}


\begin{thebibliography}{99}
\bibitem{Noord07}
\refitem{article} E.~Noordermeer and J.~M.~van~der~Hulst, \mnras\
{\bf 376}, 1480 (2007).

\bibitem{Boselli06}
\refitem{article} A.~Boselli and G.~Gavazzi,  \pasp\ {\bf 118},
517 (2006).

\bibitem{Boselli09}
\refitem{article} A.~Boselli, S.~Boissier, L~Cortese, and
G.~Gavazzi, Astron. Nachr. {\bf 330},  904 (2009).

\bibitem{Barr07}
\refitem{article} J.~M.~Barr, A.~G.~Bedregal,
A.~Arag\'on-Salamanca, et al., \aaa\ {\bf 470}, 173 (2007).

\bibitem{Aragon07}
\refitem{article} A.~Arag\'on-Salamanca, IAUS {\bf 245}, 285
(2008).

\bibitem{Bekki11}
\refitem{article} K.~Bekki and W.~J.~Couch, \mnras\ {\bf 415},
1783 (2011).

\bibitem{vdBergh09}
\refitem{article} S.~van~den~Bergh, \apj\ {\bf 694}, L120 (2009).

\bibitem{Williams10}
\refitem{article} M.~J.~Williams, M.~Bureau, and M.~Cappellari,
\mnras\ {\bf 409}, 1330 (2010).

\bibitem{Sil-Chil11}
\refitem{article} O.~K.~Sil'chenko and I.~V.~Chilingarian, Astron.
Lett.\ {\bf 37}, 1 (2011).

\bibitem{Noord08}
\refitem{article} E.~Noordermeer, M.~R.~Merrifield, L.~Coccato, et
al., \mnras\ {\bf 384}, 943, (2008).

\bibitem{Bournaud05}
\refitem{article} F.~Bournaud, C.~J~Jog, and F.~Combes, \aaa\ {\bf
437}, 69 (2005).

\bibitem{ZKhS11}
\refitem{article} A.~V.~Zasov, A.~V.~Khoperskov, and
A.~S.~Saburova, Astron. Lett. {\bf 37}, 374 (2011).

\bibitem{Cortesi11}
\refitem{article} A.~Cortesi, M.~R.~Merrifield, M.~Arnaboldi, et
al., \mnras\ {\bf 414}, 642 (2011).

\bibitem{FKh11}
\refitem{book}
A.~M.~Fridman and A.~V.~Khoperskov, {\it The Physics of Galactic
Disks} (Fizmatlit, Moscow, 2011) [in Russian].

\bibitem{Griv12}
\refitem{article} E.~Griv and M.~Gedalin,  \mnras\ {\bf 422}, 600
(2012).

\bibitem{KhZT03}
\refitem{article} A.~V.~Khoperskov, A.~V.~Zasov, and
N.~V.~Tyurina, Astron. Rep. {\bf 47}, 357 (2003).

\bibitem{ZKhT04}
\refitem{article} A.~V.~Zasov, A.~V.~Khoperskov, and
N.~V.~Tyurina, Astron. Lett. {\bf 30}, 593 (2004).

\bibitem{Bottema88}
\refitem{article} R.~Bottema, \aaa\ {\bf 197}, 105 (1988).

\bibitem{Bottema93}
\refitem{article} R.~Bottema, \aaa\ {\bf 275}, 16 (1993).

\bibitem{Z-Sab12}
\refitem{article} A.~V.~Zasov and A.~S.~Saburova, Astron. Lett.
(in press).

\bibitem{Sab11}
\refitem{article} A.~S.~Saburova, Astron. Rep.  {\bf 55}, 409
(2011).

\bibitem{ZMKhS08}
\refitem{article} A.~V.~Zasov, A.~V.~Moiseev, A.~V.~Khoperskov,
and E.~A.~Sidorova, Astron. Rep. {\bf 52}, 79 (2008).

\bibitem{ChNCC10}
\refitem{article} I.~V.~Chilingarian, A.~P.~Novikova, V.~Cayatte,
et al.,  \aaa\ {\bf 504}, 389 (2009).

\bibitem{Noord05}
\refitem{article} E.~Noordermeer, J.~M.~van~der~Hulst, R.~Sancisi,
et al., \aaa\ {\bf 442}, 137 (2005).

\bibitem{Noord_ea07}
\refitem{article} E.~Noordermeer, J.~M.~van~der~Hulst, R.~Sancisi,
et al., \mnras\ {\bf 376}, 1513 (2007).

\bibitem{Noord_Verh07}
\refitem{article} E.~Noordermeer and M.~A.~W.~Verheijen, \mnras\
{\bf 381}, 1463 (2007).

\bibitem{Ho_Filippenko97}
\refitem{article} L.~C.~Ho, A.~V.~Filippenko, W.~L.~Sargent, and
C.~Y.~Peng, \apjs\ {\bf 112}, 391 (1997).

\bibitem{Mendez08}
\refitem{article} J.~Mendez-Abreu, J.~A.~L.~Aguerri,
E.~M.~Corsini, and E.~Simonneau, \aaa\ {\bf 478}, 353 (2008).

\bibitem{Simien98}
\refitem{article} F.~Simien and Ph.~Prugniel, \aas\ {\bf 131}, 287
(1998).

\bibitem{AfMois05}
\refitem{article} V.~L.~Afanasiev and A.~V.~Moiseev, Astron. Lett.
{\bf 31}, 194, (2005).

\bibitem{KatkovChil11}
\refitem{misc} I.~Y.~Katkov and I.~V~Chilingarian, ASP Conf. Proc.
{\bf 442}, 143 (2011).

\bibitem{Koleva2009ulyss}
\refitem{article} M.~Koleva, Ph.~Prugniel, A.~Bouchard, and Y.~Wu,
\aaa\ {\bf 501}, 1269 (2009).

\bibitem{Koleva2008}
\refitem{article} M.~Koleva, Ph.~Prugniel, and S.~De~Rijcke,
\mnras\ {\bf 385}, 1998 (2008).

\bibitem{van_der_Marel1993}
\refitem{article} R.~P.~van~der~Marel and M.~Franx, \apj\ {\bf
407}, 525 (1993).

\bibitem{LaBorgne2004}
\refitem{article} D.~Le~Borgne, B.~Rocca-Volmerange, Ph.~Prugniel,
et al., \aaa\ {\bf 425}, 881 (2004).

\bibitem{CappellariEmsellem2004}
\refitem{article} M.~Cappellari and E.~Emsellem, \pasp\ {\bf 116},
138 (2004).

\bibitem{Chilingarian2007}
\refitem{misc} I.~Chilingarian, Ph.~Prugniel, O.~Sil'chenko, and
M.~Koleva, IAUS {\bf 241}, 175 (2007).

\bibitem{sil12a}
\refitem{misc} O.~Sil'chenko, IAUS {\bf 284} (in press);
astro-ph:1112.3771.

\bibitem{sil12b}
\refitem{article} O.~K.~Sil'chenko, I.~S.~Proshina, A.~P.~Shulga,
and S.~E.~Koposov, submitted to \mnras.

\bibitem{Silk09}
\refitem{article} N.~Silk,  \apj\ {\bf 700}, 262 (2009).

\end{thebibliography}
\end{document}